\documentclass{ws-p9-75x6-50}

\begin{document}

\title{Recent results by the MOA group on gravitational microlensing}

\author{P. Yock$^1$, I. Bond$^{1,2}$, N. Rattenbury$^1$, J. Skuljan$^2$, T. Sumi$^3$, F. Abe$^3$, R. Dodd$^{1,4}$, J. Hearnshaw$^2$, M. Honda$^5$, J. Jugaku$^6$, P. Kilmartin$^2$, A. Marles$^1$, K. Masuda$^3$, Y. Matsubara$^3$, Y. Muraki$^3$, T. Nakamura$^7$, G. Nankivell$^4$, S. Noda$^3$, C. Noguchi$^3$, K. Ohnishi$^8$, M. Reid$^9$, To. Saito$^{10}$, H. Sato$^7$, M. Sekiguchi$^5$, D. Sullivan$^9$, M. Takeuti$^{11}$, Y. Watase$^{12}$ and T. Yanagisawa$^3$}

\address{$^1$University of Auckland, Auckland, NZ; $^2$Canterbury University, Christchurch, NZ; $^3$Nagoya University, Nagoya 464, Japan; $^4$Carter Observatory, Wellington, NZ; $^5$Institute for Cosmic Ray Research, University of Tokyo, Tokyo 188, Japan;$^6$Research Institute of Civilization, Tama 206, Japan; $^7$Kyoto University, Kyoto 606, Japan; $^8$Nagano National College of Technology, Japan; $^9$Victoria University, Wellington, NZ; $^{10}$Tokyo Metropolitan College of Aeronautics, Tokyo 116, Japan; $^{11}$Tohoku University, Sendai, Japan; $^{12}$KEK Laboratory, Tsukuba 305, Japan\\E-mail: moa-group@vuw.ac.nz}


\maketitle

\abstracts{Recent work by the MOA gravitational microlensing group is briefly described, including (i) the current observing strategy, (ii) use of a high-speed parallel computer for analysis of results by inverse ray shooting, (iii) analysis of the light curve of event OGLE-2000-BUL-12 in terms of extra-solar planets, and (iv) the MOA alert system using difference imaging.}

\section{Observing strategy of MOA}
The gravitational microlensing technique$^{1,2}$ is being used for a variety of purposes by a Japan/NZ group called MOA. A 60-cm wide-field telescope at the Mt John Observatory in New Zealand and a large format CCD camera that was supplied by the National Astronomical Observatory of Japan are used$^3$. The small aperture and wide field of the telescope favour the study of microlensing events in which the magnification is high. Previous experience has shown that the peaks of these events include significant information on the locations\footnote{Knowledge of the locations of lenses is useful for determining the fraction of dark matter composed of 'machos'. See, e.g., Sumi and Honma$^4$.} of lens-stars$^5$, on the sizes of source-stars$^5$, on the atmospheres of source-stars$^6$, and on the planetary systems of lens-stars$^{7,8}$. Approximately 20 square degrees in both the galactic bulge and the LMC are observed a few times per night, weather permitting. The bulge data are analysed on-line (see section 4 below) and alerts issued of high magnification events for follow-up by other groups. Also, some alerts issued by other groups are followed. For some events the baseline magnitude is below the limiting magnitude of MOA. Baseline data are required from other groups with larger telescopes, or from the Hubble Space Telescope, for these events.       
\section{Inverse ray shooting with the Kalaka parallel computer}
Data for individual events are analysed using the inverse ray shooting technique of Wambsganss$^9$. This includes the effects of multi-component lenses and  limb-darkened sources of finite size, but it requires considerable computing power. The parallel computer 'Kalaka' at the University of Auckland is used for this work  \footnote{http://www.scitec.auckland.ac.nz/$^{\sim}$peter/kalaka.html}. This has 200 processors. The full parameter space for a binary lens and a finite-sized source star can be compared with a given dataset in a few hours. For a triple lens a few months is required.
\begin{figure}[t]
\begin{center}
\epsfxsize=16pc % will enlarge or reduce the postscript figures based on the xsize
\epsfbox{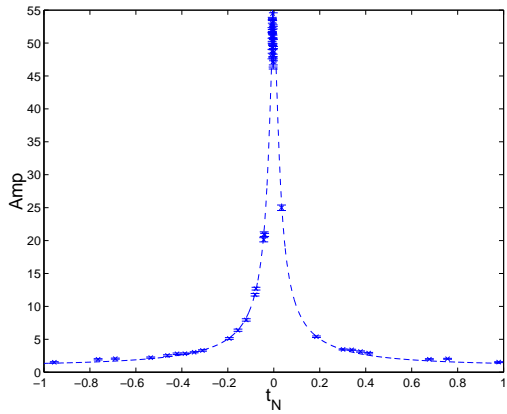} % postscript image file name
\caption{Single-lens fit to the light curve for OGLE-2000-BUL-12.\label{fig:radish}}
\end{center}
\end{figure}
\section{Test analysis of OGLE-2000-BUL-12}
An event with magnification $\sim50$ that was reported recently by the OGLE group, OGLE-2000-BUL-12, has been analysed using the Kalaka computer. Data for the event were obtained from the early warning alert site of the OGLE group\footnote{http://www.astrouw.edu.pl/$^{\sim}$ftp/ogle/ogle2/ews/ews.html}. These data were derived from the 'fixed position' version of DoPHOT$^{10}$, and they do not allow for possible blending of the source star. The present analysis is therefore preliminary. The light curve for the event is shown above in Fig. 1 together with a best single-lens fit. It was found that better fits could be obtained with two classes of binary lenses, corresponding to a lens star with either a low-mass or a high-mass planet, respectively. These are described below.        
\subsection{Low-mass planet solution}
Reasonable fits to the data were obtained with a planet at a projected separation from the lens star of $(0.9-1.1)R_E$ and mass fraction $(1-4)\times10^{-5}$, as shown in Fig. 2. For a lens mass $\sim0.3M_{\odot}$, and lens and source distances $\sim$ 7 and 9 kpc respectively, these solutions correspond to a planet with mass $\sim(1-4)M_E$ at a projected separation of $\sim2$ AU. A non-limb-darkened source star with radius $1R_{\odot}$ was assumed in the fitting.       
\begin{figure}[t]
\begin{center}
\epsfxsize=19pc % will enlarge or reduce the postscript figures based on the xsize
\epsfbox{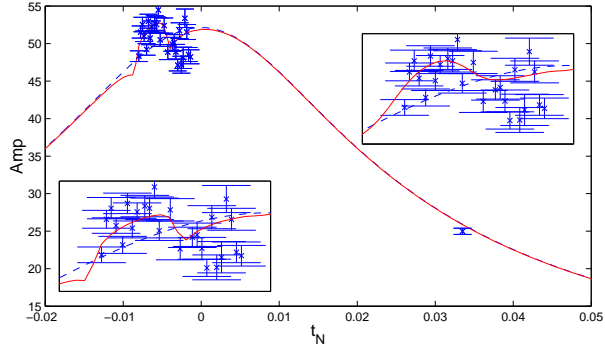} % postscript image file name
\caption{Low-mass planet fit to OGLE-2000-BUL-12. The projected coordinates and mass-fraction of the planet are $(x_p,y_p,q_p)=(-0.32,-0.96,1.6\times10^{-5})$. The source-star moves horizontally left-to-right through the origin. The lens-star is at $(0,u_{min})=(0,0.0192)$. The dashed curve is the single-lens fit from figure 1. The lower and upper insets show fits with $(x_p,y_p,q_p)=(-0.32,-0.95,10^{-5})$ and $(-0.37,-1.03,4.1\times10^{-5})$ respectively.\label{fig:radish}}
\end{center}
\end{figure}
\subsection{High-mass planet solution}
Reasonable fits to the data were also obtained with a planet at a projected separation of $(0.5-0.6)R_E$ or $(1.6-2.0)R_E$ and mass fraction $(3-6)\times10^{-3}$, as shown in Fig.3. The two distinct solutions, one with the planet inside the Einstein radius and one with it outside, are as expected$^{11}$. The solutions correspond to a planet with mass $\sim(1-2)M_J$ at $\sim1$ AU or $\sim(3-4)$ AU.  
\begin{figure}[t]
\begin{center}
\epsfxsize=19pc % will enlarge or reduce the postscript figures based on the xsize
\epsfbox{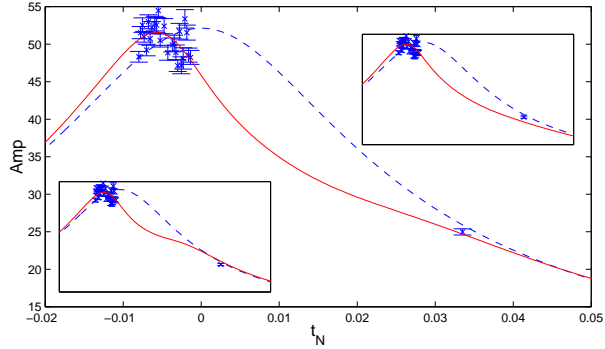} % postscript image file name
\caption{High-mass planet fit to OGLE-2000-BUL-12 with $(x_p,y_p,q_p)=(-0.31,0.52,4.8\times10^{-3})$.The insets show fits for $(x_p,y_p,q_p)=(-0.26,0.59,3.0\times10^{-3})$ and $(-0.34,0.46,6.2\times10^{-3})$.\label{fig:radish}}
\end{center}
\end{figure}
\subsection{Discussion}
As noted above, the above analysis is only preliminary, because blending has not been allowed for. Additional solutions with a planet very close to the Einstein radius or with more than one planet may also be possible. Further data for the event, if available, should differentiate between the possible solutions. Further data may also become available in the future that would pinpoint the planetary parameters. In approximately one decade, the lens and source stars may diverge sufficiently to enable the next generation of space telescopes to resolve them separately. In this case, the mass of the lens star, and the distances of the lens and source stars, may be directly measurable. This would leave no parameters undetermined.  
\section{MOA alert system using difference imaging}
On-line data analysis has recently been implemented by MOA. This uses difference imaging in a manner that is similar to that of Alard and Lupton$^{12}$. This enables events that are magnified from beneath the limiting magnitude to be readily detected, and also it ameliorates the effects of crowding and blending that are necessarily present. Some details of the technique were described previously$^{13}$. Full details will be presented in the future. Alerts of events in progress are being reported on the web\footnote{http://www.phys.canterbury.ac.nz/$^{\sim}$physib/alert/alert.html}.
\section*{Acknowledgments}
The authors thank the OGLE group for making their data available and for discussion, and Peter Dobcsanyi and Michael Harre for assistance. Financial support by the Marsden Fund of NZ and the conference organizers is gratefully acknowledged.

\end{document}